\documentclass[aps,pra,twocolumn,superscriptaddress,preprintnumbers,nofootinbib,longbibliography,floatfix]{revtex4-2}
\usepackage{graphicx}
\usepackage{dcolumn}
\usepackage{bm}
\usepackage{enumitem}
\usepackage{numprint}
\usepackage{mathtools}
\usepackage{amsfonts}
\usepackage{siunitx}
\usepackage{adjustbox}
\usepackage{float}
\restylefloat{table}
\usepackage{subcaption}
\usepackage{multirow}
\usepackage[normalem]{ulem}
\usepackage{tikz}
\usetikzlibrary{positioning, shapes, arrows.meta}
\usepackage{quantikz} % For quantum circuit diagrams
\usepackage{graphicx} % For controlling figure size
\usepackage{physics}

\usepackage[colorlinks=true
,urlcolor=blue
,anchorcolor=blue
,citecolor=blue
,filecolor=blue
,linkcolor=blue
,menucolor=blue
,pagecolor=blue
]{hyperref}
\usepackage{braket}
\usepackage{amsmath,amsfonts}

\def\begsub#1#2\endsub{\begin{subequations}\label{eq:#1}\begin{align}#2\end{align}\end{subequations}}
\def\ba#1\ea{\begin{align}#1\end{align}}
\def\bas#1\eas{\begin{align*}#1\end{align*}}

\usepackage{acronym}

\begin{document}
\npthousandsep{\ensuremath\,\allowbreak}
%%%%%%%%%%%%
%
%  Header information
%
%%%%%%%%%%%%

\title{Towards Designing Scalable Quantum-Enhanced Generative Networks for Neutrino Physics Experiments with Liquid Argon Time Projection Chambers}% Force line breaks with \\
\thanks{This manuscript has been authored by UT-Battelle, LLC, under Contract No. DE-AC0500OR22725 with the U.S. Department of Energy. The United States Government retains and the publisher, by accepting the article for publication, acknowledges that the United States Government retains a non-exclusive, paid-up, irrevocable, world-wide license to publish or reproduce the published form of this manuscript, or allow others to do so, for the United States Government purposes. The Department of Energy will provide public access to these results of federally sponsored research in accordance with the DOE Public Access Plan.}%

\author{Andrea Delgado}
\email{delgadoa@ornl.gov}
\affiliation{Physics Division, Oak Ridge National Laboratory, Oak Ridge, TN 37831}

\author{Diego Venegas-Vargas}
\email{dvenega1@jh.edu}
\affiliation{Department of Physics and Astronomy, Johns Hopkins University, Baltimore, MD 21218}

\author{Adam Huynh}
\affiliation{Powell High School, Powell, TN 37849}

\author{Kevon Carroll}
\affiliation{Haslam College of Business, University of Tennessee, Knoxville, TN 39716}

\begin{abstract}
Generative modeling for high-resolution images in Liquid Argon Time Projection Chambers (LArTPC), used in neutrino physics experiments, presents significant challenges due to the complexity and sparsity of the data. This work explores the application of quantum-enhanced generative networks to address these challenges, focusing on the scaling of models to handle larger image sizes and avoid the often encountered problem of mode collapse. To counteract mode collapse, regularization methods were introduced and proved to be successful on small-scale images, demonstrating improvements in stabilizing the training process. Although mode collapse persisted in higher-resolution settings, the introduction of these techniques significantly enhanced the model's performance in lower-dimensional cases, providing a strong foundation for further exploration. These findings highlight the potential for quantum-enhanced generative models in LArTPC data generation and offer valuable insights for the future development of scalable hybrid quantum-classical solutions in nuclear and high-energy physics.
\end{abstract}

\maketitle

%\tableofcontents

%%%%%%%%%%%%
%
%  Introduction
%
%%%%%%%%%%%%
\section{Introduction}
\label{sec:intro}

%\begin{itemize}
%    \item Brief overview of hybrid quantum-classical models and their potential in machine learning.
%    \item Motivation for exploring large-scale utility and challenges in modeling complex data.
%    \item Motivation for using a normalizing flow model with quantum-classical coupling.
%    \item Brief overview of the model proposed in Ref.~\cite{Zhang:2024sfa}. Talk about how it compares to Ref.~\cite{Rousselot_2024}
%    \item \textbf{Objective of the manuscript: } To investigate the performance of the hybrid model on real-world data and understand its limitations and robustness.
%\end{itemize}

Liquid Argon Time Projection Chamber (LArTPC) experiments play a crucial role in nuclear and high-energy physics, including neutrino detection~\cite{Balasubramanian:2023pap,Abratenko2024} and the search for rare processes such as neutrinoless double beta decay~\cite{PhysRevD.106.092002}. These detectors provide detailed 2D and 3D reconstructions of particle interactions by capturing the ionization charge produced as particles traverse liquid argon. Although LArTPC data offers valuable insights into fundamental physics, their complex, high-dimensional nature poses significant challenges for reconstruction, especially in sparse signal regions and noisy environments.

The complexity of reconstructing particle trajectories from LArTPC images has led to the exploration of alternative methods, such as generative modeling techniques, to simulate or augment data. Generative models are well-suited for managing the complexity of LArTPC datasets by efficiently learning their underlying data distributions. However, training traditional models like Generative Adversarial Networks (GANs) or Variational Autoencoders (VAEs) to generate realistic LArTPC data is both computationally expensive and challenging, especially as datasets grow in size and complexity. These models often struggle to accurately capture the fine-grained details of particle interactions, and their high computational cost limits scalability~\cite{PhysRevD.109.072011,Lutkus2022}.

Hybrid quantum-classical models represent a promising frontier in machine learning by combining the strengths of classical neural networks with the unique capabilities of quantum computing. Quantum generative models have shown potential in various nuclear and particle physics tasks, such as learning probability distributions for data generation~\cite{PhysRevD.106.096006,delgado2023identifyingoverparameterizationquantumcircuit} and performing anomaly detection~\cite{Bermot_2023}. These models typically involve embedding quantum circuits within a classical framework, where the quantum component is responsible for encoding or generating data representations in a higher-dimensional Hilbert space. The parameters of the quantum circuits are trained using classical methods, as seen in variational quantum algorithms~\cite{Cerezo_2021}. Classical networks, such as feed-forward or convolutional neural networks, are often used to process or enhance the quantum outputs~\cite{zaman2024comparativeanalysishybridquantumclassical}. This hybrid coupling allows models to exploit the computational advantages of quantum systems while maintaining the flexibility and efficiency of classical training. Consequently, hybrid models can improve performance on tasks that require complex, high-dimensional computations. As quantum hardware advances, these models are increasingly explored for their potential to solve problems that are intractable using classical methods alone.

However, deploying these models in real-world scenarios poses significant challenges. Complex data, particularly from high-dimensional or noisy environments, requires models that can efficiently learn and generalize from diverse datasets~\cite{Wang2023,Jahin2023}. The scaling of hybrid quantum-classical models to large, complex datasets remains an open research question~\cite{Guarasci2022,Peixoto2023}. Addressing this challenge necessitates exploring models capable of capturing intricate distributions, while also being robust to hardware noise, a prevalent issue in current quantum devices~\cite{PRXQuantum.5.030314,Rehm2023}.

In addition, training quantum circuits in conjunction with classical networks introduces several unique challenges. One of the primary issues is the inefficiency of quantum gradient computation, which requires specialized techniques like parameter-shift rules or finite-difference methods~\cite{PhysRevA.99.032331}. These methods are often computationally expensive, particularly when scaling to larger circuits or deeper networks. Furthermore, noise from quantum hardware --- whether due to decoherence, gate errors, or limited qubit connectivity --- further complicates the training process, as the classical network must learn to compensate for the imperfections introduced by the quantum component.

In this context, normalizing flow models~\cite{rezende2016variationalinferencenormalizingflows} (see Ref.~\cite{Kobyzev_2021} for a comprehensive review), which allow for tractable density estimation and flexible generative modeling, provide a natural framework for quantum-classical coupling. By integrating quantum computing into the flow-based architecture, it becomes possible to leverage quantum processing for tasks such as sampling and optimization, which are computationally expensive for classical models~\cite{Cranmer2023,Coyle2020}. This coupling has the potential to enhance the model's ability to learn complex data distributions while maintaining scalability. This manuscript builds on the models proposed in Refs.~\cite{Zhang_2024sfa,Rousselot_2024}, which introduce \emph{hybrid quantum-classical normalizing flows} (HQCNF) for generative modeling. Our work investigates the performance of the hybrid model in image generation tasks using LArTPC data, with a particular focus on its robustness and potential advantages in handling complex datasets.

\section{Model Architecture}
\label{sec:model}

%\begin{itemize}
%    \item \sout{Detailed description of the hybrid normalizing flow model: Structure of the two classical networks and their functions. Integration of the quantum circuit, choice of quantum gates, qubits, and the entanglement strategy.}
%    \item \sout{Mathematical formulation of the normalizing flow and its implementation.}
%    \item Explanation of the coupling mechanism between the classical networks and the quantum circuit.
%    \item Training procedure and loss functions used for the model.
%\end{itemize}

Normalizing flows are a powerful technique in probabilistic modeling for transforming a simple, tractable probability distribution into a more complex one while retaining the ability to compute the likelihood of data points efficiently ~\cite{rezende2016variationalinferencenormalizingflows}. The central idea is to apply a sequence of invertible transformations to a simple base distribution, such as a Gaussian, to model a complex target distribution.

A normalizing flow begins with a simple base distribution, typically a multivariate Gaussian. Let \( z_0 \) be a random variable drawn from this base distribution, \( z_0 \sim p_{z_0}(z_0) \), where \( p_{z_0}(z_0) = \mathcal{N}(z_0 \mid 0, I) \), representing a Gaussian with mean zero and identity covariance matrix. A normalizing flow consists of a sequence of invertible, differentiable transformations $f_1, f_2, \dots, f_K$. Each transformation $f_{i}$ maps one random variable to another. Starting from $z_{0}$, each transformation generates a new random variable $z_{i}$ as follows:

\begin{equation}
    z_i = f_i(z_{i-1}), \quad z_0 \sim p_{z_0}(z_0)
\end{equation}

Since each transformation is invertible, we can express the inverse transformation as $z_{i-1}=f_{i}^{-1}(z_{i})$. To compute the probability density of the final variable $z_K$, we can apply the change of variables formula. The probability density of $z_K$, denoted as $p_{z_K}(z_K)$, is related to the base distribution $p_{z_0}(z_0)$ as follows:

\begin{equation}
p_{z_K}(z_K) = p_{z_0}(z_0) \prod_{i=1}^{K} \left| \det \frac{\partial f_i(z_{i-1})}{\partial z_{i-1}} \right|^{-1}
\end{equation}
after applying $K$ transformations. Since the Jacobian of the inverse transformation $f_{i}^{-1}$ is the inverse of the Jacobian of the forward transformation $f_{i}$, we get this simplified version.

The transformations $f_{i}$ are typically parameterized by neural networks, and the parameters are optimized to maximize the likelihood of the observed data under the final distribution $p_{z_K}(x)$. The training objective is to maximize the log-likelihood of the data $x$:

\begin{equation}
    \log p_{z_K}(x) = \log p_{z_0}(f^{-1}(x)) - \sum_{i=1}^{K} \log \left| \det \frac{\partial f_i(z_{i-1})}{\partial z_{i-1}} \right|
\end{equation}
where $f^{-1}(x)$ represents the inverse transformation that maps the data back to the base distribution

Once trained, generating samples from the model involves drawing $z_{0}$ from the base distribution and then applying the sequence of transformations:

\begin{equation}
    x = f_K \circ f_{K-1} \circ \dots \circ f_1 (z_0)
\end{equation}
allowing to sample from the complex distribution learned by the normalizing flow.

The HQCNF model is designed to leverage both classical and quantum computing paradigms to perform generative modeling tasks. The model combines multiple layers of classical neural networks with a quantum circuit to create a powerful hybrid architecture capable of learning complex data distributions.

In generative modeling, the classical network often learns to correct for quantum noise or amplify quantum-enhanced features that contribute to the generation process. In a normalizing flow context, quantum circuits can be employed as part of the flow transformations, potentially allowing the model to better approximate complex data distributions.

The classical part of the model consists of a series of normalizing flow transformations, each designed to apply an invertible affine transformation to input data. These transformations are parameterized by neural networks that learn the scale and translation functions required for the bijective mappings.

The model is built using a sequence of \emph{affine coupling layers}, where each layer is responsible for transforming the input data in both the forward and reverse directions. Each coupling layer consists of two components: a scaling function($s(\cdot)$) and a translation function ($t(\cdot)$), both of which are implemented as fully connected neural networks. These neural networks have three layers, and their architecture gradually reduces the dimensionality from the input to the output size. Nonlinearity is introduced using the $\tanh$ activation function, which helps capture complex dependencies in the data.

\emph{Network structure:} Each neural network within a coupling layer receives the input data of size $n$, splits it in half to produce representations of size $n/2$, and then gradually increases the dimensionality back to the original size. This structure allows the model to progressively learn complex transformations while maintaining computational efficiency.

\emph{Forward and inverse transformation:} During the forward pass (inference), the model applies an affine transformation to the input data. This transformation is governed by the learned scale and translation functions $s(\cdot)$ and $t(\cdot)$, defined as:

\begin{equation}
z_i = x_i \odot \exp(s(x_{<i})) + t(x_{<i})
\end{equation}
where $\odot$ represents element-wise multiplication, and $z_i$ is the transformed output. The inverse transformation is computed by reversing these operations, allowing the model to establish a bijective mapping between the input data and the latent space. This property ensures that the model is invertible and can generate samples from the target distribution.

Additionally, the \emph{log-determinant of the Jacobian} is computed for each transformation during both forward and inverse passes. This term is critical for maximizing the likelihood during training, as it ensures the proper adjustment of the model's density under the change of variables framework.

\emph{Layer composition:} Multiple affine coupling layers are stacked sequentially to construct the full normalizing flow model. The stacking allows the model to learn increasingly complex transformations of the data, enhancing its ability to approximate more intricate distributions while preserving computational tractability and invertibility. %Each layer contributes to the overall transformation, making the normalizing flow a powerful tool for flexible density estimation.

The quantum part of the HQCNF model consists of a parameterized quantum circuit designed to enhance the expressiveness of the generative model by leveraging quantum entanglement and superposition. The input data to the quantum circuit is first encoded using \emph{amplitude embedding}. The quantum circuit is constructed using a sequence of parameterized rotation gates (RY) and entangling gates (CNOTs). For each layer in the quantum circuit, the rotation gates are applied to each qubit with a set of learnable parameters. These parameters are obtained from the outputs of the classical neural networks. CNOT gates are used to entangle adjacent qubits, creating correlations between the qubits that enhance the circuit's ability to represent complex probability distributions.

%----------------------------------------------------------
%\begin{figure}[h]
%\centering
%\includegraphics[]{}
%\caption{Quantum circuit structure for the hybrid quantum-classical normalizing flow model. The circuit consists of multiple layers of parameterized rotation gates (RY) and entangling gates (CNOT) to create a highly expressive quantum model.}
%\end{figure}
%----------------------------------------------------------

The loss function for the hybrid quantum-classical normalizing flow model is given by
\begin{equation}
L(\theta) = - \sum_{i=1}^{m} \log \pi(f^{-1}_{\theta}(x^{(i)})) + \sum_{i=1}^{m} \log \left| \det J^{(i)}_{\theta} \right|,
\label{eq:loss}
\end{equation}

\noindent where $(\pi(f^{-1}_{\theta}(x)))$ represents the prior distribution and $(J^{(i)}_{\theta})$ is the Jacobian matrix of the transformation $(f_{\theta})$ applied to the input data.

%The loss function for the hybrid quantum-classical normalizing flow model is derived from the classical normalizing flow framework, adapted for the quantum setting as defined in Eq.~\ref{eq:loss} in Section~\ref{sec:model}. In this expression, $\pi(f^{-1}_{\theta}(x^{(i)}))$ represents the prior distribution, which is typically assumed to be a standard Gaussian or another chosen distribution. The Jacobian matrix $(J^{(i)}_{\theta})$ of the transformation $f_{\theta}$ applied to the input data plays a critical role in ensuring that the transformation is bijective and preserves probability mass \diego{both $\pi(f^{-1}_{\theta}(x^{(i)}))$ and $(J^{(i)}_{\theta})$ were defined in the previous paragraph, so this is a bit repetitive. }.
The loss function has two main components: the first term measures the discrepancy between the transformed data and the target distribution, while the second term accounts for the invertibility and complexity of the transformation. Together, these terms aim to minimize the difference between the distribution of generated data and the real data distribution, while maintaining the expressiveness of the model through the Jacobian determinant (see Appendix~\ref{sec:jacobian}).

%-----------------------------------------------------------
\section{Experiments}
\label{sec:experiments}

%\begin{itemize}
%    \item \sout{Dataset overview: description of the MNIST dataset and the real world dataset.}
%    \item Training details: pre-processing, model initialization, hyperparameters, computational resources, and training duration.
%    \item \sout{Description of the evaluation metrics:fidelity, loss functions, computational complexity, etc.}
%    \item Results on the real world dataset: challenges faces, performance metrics, and visualizations.
%\end{itemize}

%Projections in the XZ-view of LArTPC images are also employed. 
The dataset consists of $12\times 8$ crops of particle trajectories from images found in the public \texttt{PILArNet 2D} dataset~\cite{adams2020pilarnetpublicdatasetparticle}, a repository of LArTPC-like 2D images and 3D voxelized data. The data is generated using \texttt{GEANT4}~\cite{GEANT4:2002zbu}, a simulation toolkit used to simulate the transport of particles within a volume of liquid argon. The dataset records the location of ionizations created by particles moving through the liquid argon volume.

To generate the images, the 3D ionization locations are projected onto the XY, XZ, and YZ planes. A charge diffusion model is applied during the projection to account for spatial distribution changes in both the longitudinal and transverse directions, increasing the spread of ionization energy deposits. Particles in the simulation are randomly selected from \emph{electrons, photons, muons, protons}, and \emph{charged pions}, with their starting locations uniformly distributed throughout the simulated volume.

For the images used in this study, $12\times 8$ crops are taken from the original $256\times 256$ LArTPC images, centering the crop on the brightest pixel to avoid empty regions. During data preprocessing, a common bias value of 10.0 is subtracted from all pixel values, followed by scaling the values to lie between -1 and 1. Finally, the pixel values are re-scaled to fit within the range 0 to 255, ensuring uniformity across the dataset.

%----------------------------------------------------------
% Begin figure with full page width in two-column layout
\begin{figure}[h]
\centering
\includegraphics[width=\linewidth]{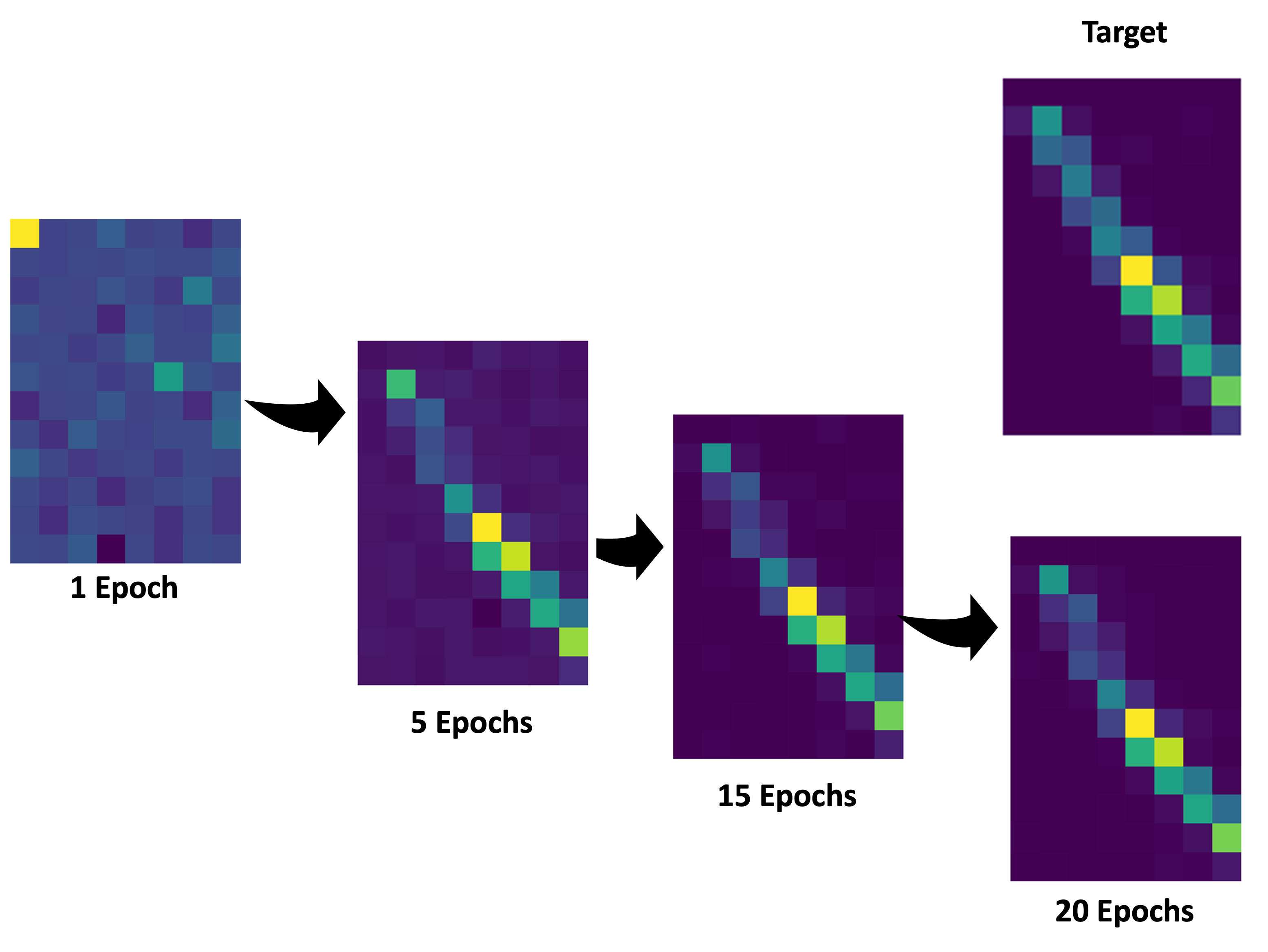}
\caption{Example training for a single image dataset. The image features a couple snapshots of generated images during training.}
\end{figure}
%----------------------------------------------------------

The computational complexity of the model is divided into two main parts: the classical neural network and the quantum circuit. The classical neural network consists of multiple sequential modules, each containing three linear layers with decreasing sizes and non-linear activation functions. The total complexity for this classical component primarily depends on the number of layers and neurons, scaling approximately as $O(n^{3})$, where $n$ is the size of the largest layer. The quantum circuit involves amplitude embedding followed by parameterized quantum gates. Its complexity is mainly determined by the number of qubits, $n_{q}$, and the depth of the circuit, resulting in a total complexity of $O(p\cdot n_{q})$, where $p$ is the number of layers. While the quantum component introduces additional complexity, it also offers the potential computational speedups for certain tasks.

The \emph{Frechet Inception Distance} (FID) score~\cite{10.5555/3295222.3295408} is used as a metric to evaluate the quality of the generated images by measuring the distance between the distribution of generated images and real images. The FID score compares the statistical properties (mean and covariance) of feature vectors extracted from a specific layer of pre-trained inception network for both real and generated images. A lower FID score indicates that the generated images are more similar to the real images, thus reflecting better model performance. Generally, an FID score close to zero is considered ideal, but achieving this score is rare and depends on the complexity of the dataset and model. The FID score is calculated using the following formula:

\begin{equation}
    \text{FID} = \left\| \mu_r - \mu_g \right\|^2 + \text{Tr} \left( \Sigma_r + \Sigma_g - 2 \left( \Sigma_r \Sigma_g \right)^{1/2} \right),
\end{equation}

\noindent where: $(\mu_r)$ and $(\mu_{g})$ are the mean feature vectors for the real and generated images, respectively. Then, $(\Sigma_r)$ and $(\Sigma_g)$ are the covariance matrices of the feature vectors for the real and generated images, respectively. Finally, $(\left\| \mu_r - \mu_g \right\|^2)$ is the squared Euclidean distance between the means of the real and generated images and $(\text{Tr}(\cdot))$ denotes the trace of a matrix.

%----------------------------------------------------------
% Begin figure with full page width in two-column layout
\begin{figure}[h]
\centering
\includegraphics[width=\linewidth]{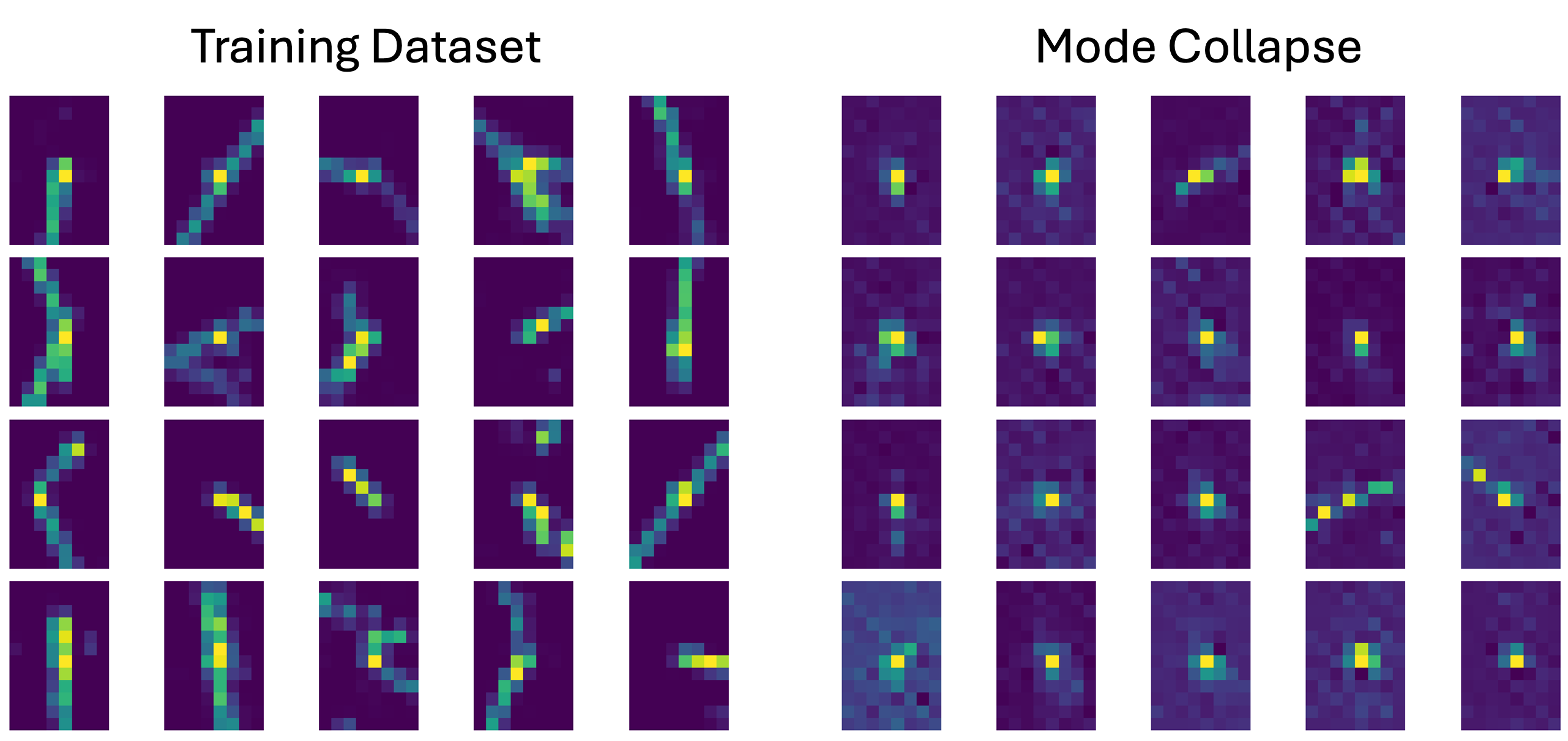}\\
\caption{Comparison of 20 instances of training images (left) vs. an instance where the model encountered mode collapse (right).}
\label{fig:modecollapse}
\end{figure}
%----------------------------------------------------------

During the initial experiment trials, mode collapse was encountered~\cite{modecollapse}. Mode collapse occurs when a generative model, instead of learning the full data distribution, produces highly repetitive outputs that represent only a limited subset of the possible variations in the training dataset. This often results in the model generating nearly identical samples, failing to capture the diversity inherent in the data. An example of mode collapse during training is shown in Fig.~\ref{fig:modecollapse}. The images on the left illustrate some samples from the training dataset. The right panel depicts several failure case where mode collapse occurred. Despite adjustments in model complexity, learning rates, and dataset size, an FID score below 25 could not be achieved. The results suggest that the model was consistently generating images with a bright cluster of pixels concentrated in the center, indicating that it had failed to capture the full distribution of the training data, instead repeatedly producing a limited and non-representative subset of features.

To address this, two strategies were explored to improve the model's performance. First, the structure of the quantum circuit ansatz was modified by using two layers of $RZRYRZ$ gates per qubit, followed by a chain of $CNOT$ gates acting on neighboring qubits. This aimed to increase the expressibility of the quantum layer, enhancing its capacity for learning complex patterns. Second, latent space regularization was introduced by adding a Variational Autoencoder (VAE)-like Kullback-Leibler (KL) divergence loss. The KL divergence encourages a smoother latent space by penalizing deviations from a standard normal distribution, thereby reducing the risk of mode collapse.

The KL-divergence term was added to the loss function as follows:

\begin{equation}
    \mathcal{L}_{KL} = -\frac{1}{2} \sum \left( 1 + \log(\sigma^2) - \mu^2 - \sigma^2 \right),
\end{equation}
where $\mu$ and $\sigma^2$ are the mean and variance of the latent distribution --- the output of networks $s(\cdot)$ and $t(\cdot)$ in our model. This term was incorporated into the total loss (Eq.~\ref{eq:loss}) alongside the classical and quantum losses to encourage the model to explore a more diverse set of latent variables and reduce overfitting to specific modes (See Appendix~\ref{sec:kldivergence} for details).

%-----------------------------------------------------------
\section{Results}
\label{sec:analysis}
A hybrid quantum-classical model was trained using the \texttt{PILArNet} dataset, resized to $12\times 8$ pixels, and focused on five different particle classes. The quantum component utilized 5 qubits, and the overall model consisted of 8 layers. The quantum loss was scaled by a multiplier of 100 to account for the different scales between the quantum and classical components. For training, independent runs were conducted, each using the Adam optimizer with either a 0.001 or 0.0001 learning rate, with the number of training epochs varying according to the specific test requirements.

\begin{figure}[h]
\centering
\includegraphics[width=\linewidth]{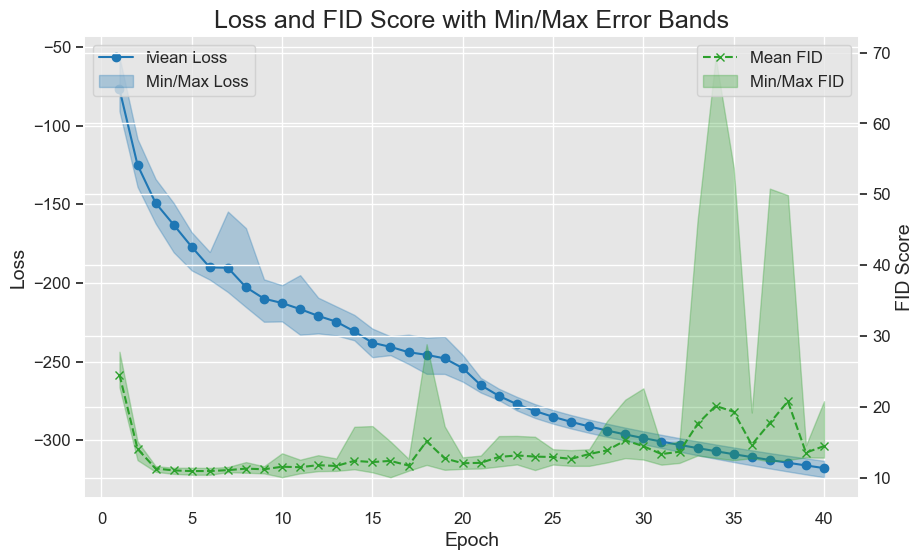}\\
\caption{The mean and the minimum-maximum range of the loss function (left) and FID score (right) across 10 different training instances of a hybrid quantum-classical model. The model was trained on the LArTPC dataset, using 5 qubits, 8 layers, and a quantum loss multiplier of 100.}
\label{fig:loss5q8l}
\end{figure}

In Fig.~\ref{fig:loss5q8l}, the loss values across 10 different runs are plotted, showing a consistent and smooth decrease throughout the training process, which indicates effective convergence. The variation between different runs was minimal, particularly during later epochs, demonstrating the model's ability to minimize the loss effectively. However, despite the smooth decline in loss, the FID score exhibited substantial variability after 10 epochs and tended to increase as training progressed. This suggests that while the internal optimization (reflected by the loss) improved steadily, the generated outputs did not fully align with the real data, as indicated by the FID score.

%----------------------------------------------------------
% Begin figure with full page width in two-column layout
%----------------------------------------------------------

%----------------------------------------------------------
% Begin figure with full page width in two-column layout
\begin{figure}[h]
\centering
\includegraphics[width=\linewidth]{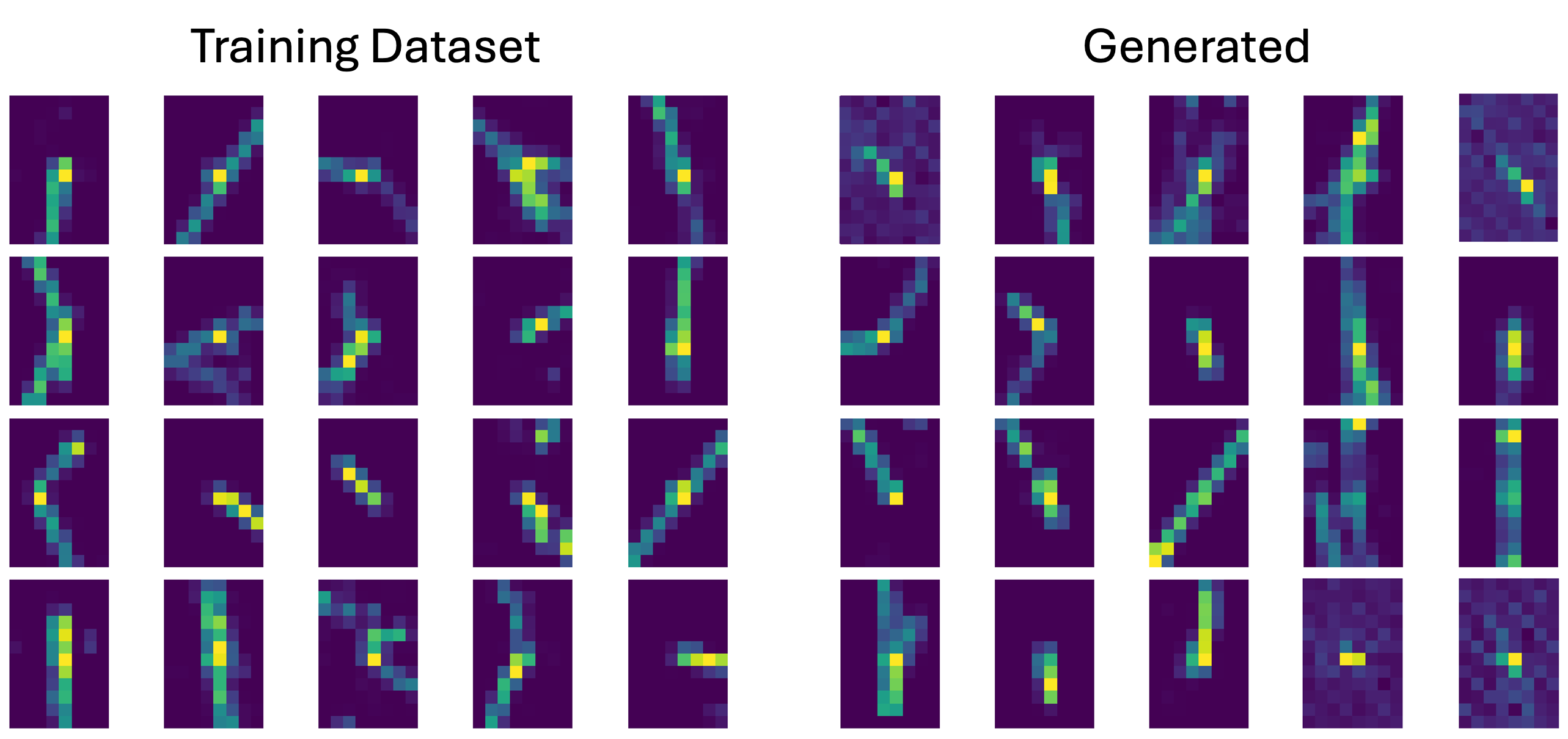}\\
\caption{Comparison of 20 images from the training dataset (left) and 20 images generated by the model (right). The generated images were produced by a model trained with 5 qubits and 8 layers. The model achieved a FID score of 9.58, indicating the quality and similarity of the generated images compared to the training data, although some cases of mode collapse can be observed.}
\label{fig:generatedimages}
\end{figure}
%----------------------------------------------------------

\subsection{Impact of dataset size on performance}
This section analyzes how the model's performance varies when trained with different dataset sizes. By gradually adjusting the number of data points, the study aims to evaluate the model's ability to generalize and learn efficiently with both limited and abundant data.

%----------------------------------------------------------
\begin{figure}[h]
\centering
\includegraphics[width=\linewidth]{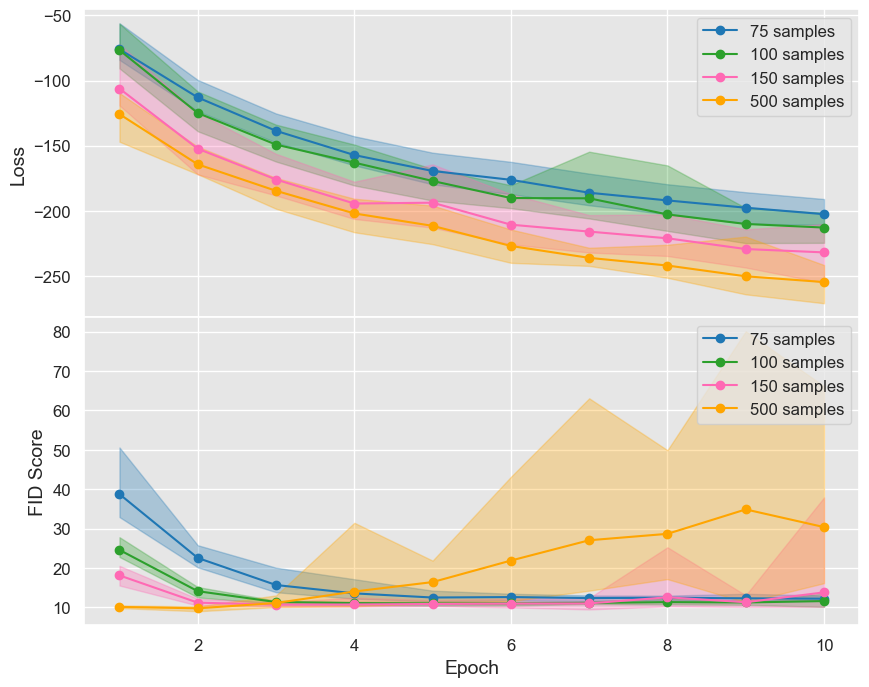}\\
\caption{The mean and the minimum-maximum range of the loss function (top) and FID score (bottom) across 10 different training instances of a hybrid quantum-classical model. The model was trained in the LArTPC dataset, using 5 qubits, 8 layers, and a quantum loss multiplier of 100.}
\label{fig:samplesplots}
\end{figure}
%----------------------------------------------------------

The plots in Fig.~\ref{fig:samplesplots} reveal distinct trends in the model's training behavior as the sample size increases. In the top plot, a clear pattern emerges: the mean loss decreases as the number of training samples increases. This trend is expected, as larger datasets generally enable the model to learn more effectively, resulting in lower loss values and more accurate updates during training. Consequently, models trained with more data tend to converge faster and reach a more optimal state, as reflected by the decreasing loss values.

In contrast, the FID score plot (bottom) presents a different outcome. For smaller sample sizes—75, 100, and 150 samples—the mean FID scores plateau at similar values after 10 training epochs, indicating comparable performance across these sizes. Notably, the spread of FID values for these smaller sample sizes remains narrow, suggesting relatively stable training outcomes. However, in the case of 500 samples, a wider spread in FID values is observed.

This increased variability in FID scores for the 500-sample case could stem from several factors. First, the model may overfit to specific features within the larger dataset, leading to more variance in performance when tested on unseen data. Additionally, a larger sample size introduces more complexity to the data distribution, which can cause variations in the optimization process, potentially leading the model to settle in different local minima during training. This may contribute to the observed spread in FID values. Moreover, training on larger datasets may be more prone to optimization instabilities, further explaining the fluctuations in FID scores.

Thus, while increasing the sample size improves the model's ability to minimize the loss function, its impact on FID scores is more nuanced. Larger datasets may introduce variability in performance metrics, highlighting the need for closer investigation.

\subsection{Effect of model depth on training}
This section explores the influence of varying number of layers in the HQCNF model on it's performance. Models were trained across 20 different instances using 100 samples in the training dataset, 5 qubits, a quantum loss multiplier of 100 and a varying number of layers.

%----------------------------------------------------------
\begin{figure}[h]
\centering
\includegraphics[width=\linewidth]{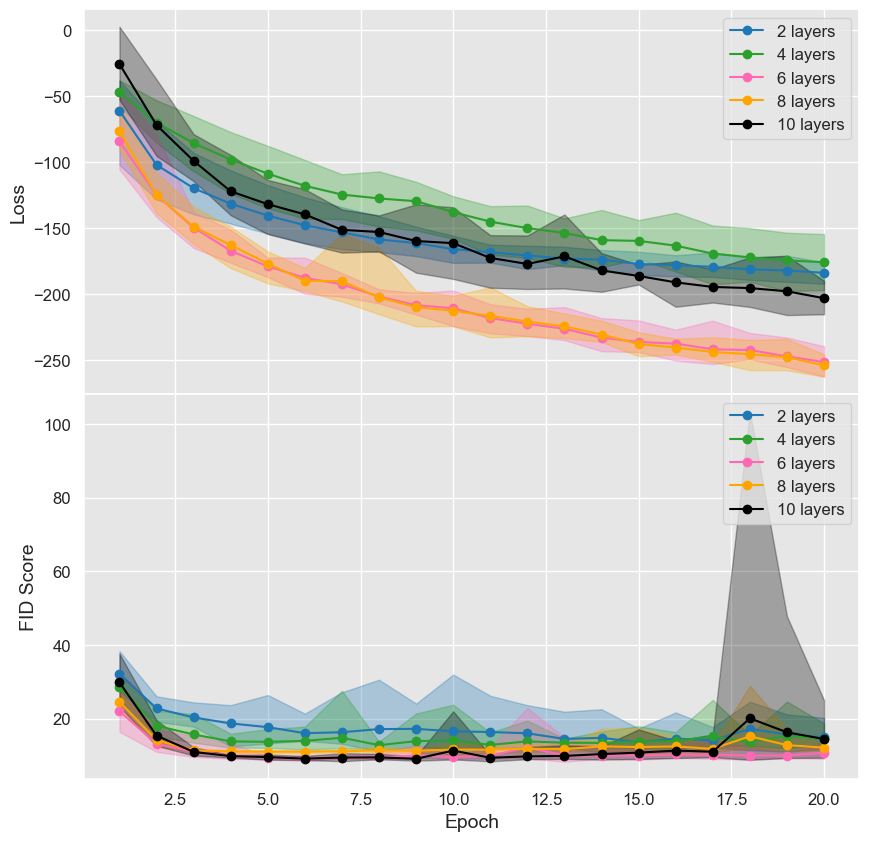}\\
\caption{The mean and the minimum-maximum range of the loss function (top) and FID score (bottom) across 20 different training instances of a hybrid quantum-classical model. The model was trained in the LArTPC dataset, using 100 samples, 5 qubits, varying number of layers, and a quantum loss multiplier of 100.}
\label{fig:layerplots}
\end{figure}
%----------------------------------------------------------

Fig.~\ref{fig:layerplots} illustrates the evolution of the loss and FID score during training as the number of layers increases from 2 to 10. The loss values show a clear decreasing trend with the increasing number of training epochs. Notably, the loss at the end of the training decreases as the number of layers increases from 2 to 8, indicating that deeper networks tend to achieve better optimization. However, the behavior for the 6 and 8-layer models is very similar, with both reaching nearly identical loss values by the end of training. In contrast, the 10-layer model exhibits a different trend, with its final loss value being comparable to the models with 2 and 4 layers, suggesting that adding more layers beyond 8 may not provide significant improvement in this case.

In the FID score, a similar overall trend is observed, with FID values generally decreasing as the number of epochs progresses, reflecting improvements in the quality of the model's output. However, a distinct feature emerges around epoch 17, where both the 8-layer and 10-layer models display a wide spread in FID values, with some runs producing significantly higher scores. This suggests that for deeper models, the training process may become less stable, leading to variations in performance across different runs. These higher FID values could be a result of the model overfitting to certain features of the data or encountering optimization difficulties, such as vanishing or exploding gradients, which are more likely to affect deeper networks. Additionally, the increased model complexity may make the training process more sensitive to initialization or other stochastic factors, further contributing to the observed variability.

This analysis suggests that while increasing the number of layers generally improves the model's ability to minimize loss and achieve lower FID scores, the training stability may decrease with very deep architectures, as seen with the 8 and 10-layer models.

%\subsection{Class-specific training performance}
%This section evaluates the model's performance when trained on individual classes of the dataset. The goal is to understand how well the model performs when focused on specific classes and whether certain classes pose more challenges than others.

\subsection{Scaling to larger image resolutions}
The model was scaled by increasing the number of qubits from 5 to 8, allowing for the processing of higher-resolution images with dimensions of $24\times 32$ pixels. This scaling aimed to capture more complexity in the data and enhance the model’s generative capabilities.

\begin{figure*}[t]
\centering
\includegraphics[width=\linewidth]{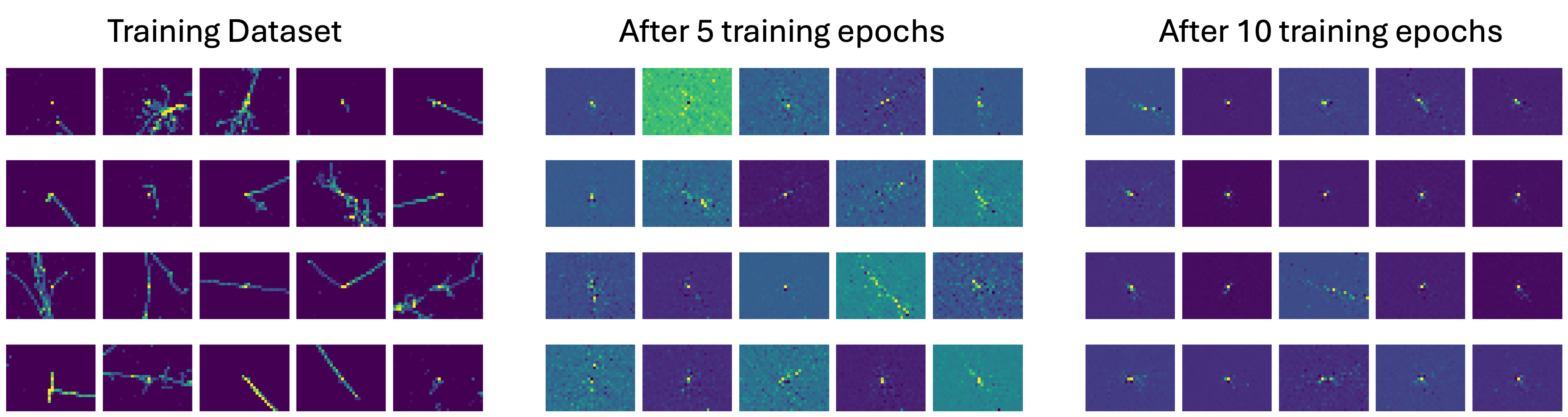}\\
\caption{Comparison of 20 `higher-resolution' (24x32 pixels) images from the training dataset (left) and generated images after 5 training epochs (middle) and 10 training epochs (right). The model initially generates bright, shower-like images, but as training progresses, it quickly falls into mode collapse, producing single-dot outputs after 10 epochs.}
\label{fig:largeplots}
\end{figure*}

Various training schemes were explored during these experiments, adjusting parameters such as the number of training steps, learning rates, and the overall size of the model. Despite these efforts, the model consistently suffered from \emph{mode collapse} (as seen in the center and right plots of Fig.~\ref{fig:largeplots}).

Even when regularization methods that had previously improved the performance of lower-dimensional models were applied, mode collapse persisted. A potential explanation for this issue lies in the nature of the higher-resolution data: as the image size increased, so did the ratio of non-zero-value pixels (see Fig.~\ref{fig:largeplots} [left]). This substantial increase in sparse pixel values may have overwhelmed the model's capacity to learn meaningful representations, causing it to collapse into producing oversimplified outputs, such as single bright dots, rather than diverse structures.

%\subsection{Robustness in the presence of hardware noise}
%This section examines how the model performs in the presence of hardware noisy, especially in quantum systems. The analysis focuses on understanding how the model's performance degrades or improves with increasing noise.

%\subsection{Evaluation using alternative metrics}
%Beyond the FID, this section presents results from additional evaluation metrics relevant to the underlying physical processes to be generated. 

%-----------------------------------------------------------
\section{Discussion}
\label{sec:discussion}
Our exploration into scaling quantum-enhanced generative models for LArTPC has provided valuable insights into the challenges and potential of hybrid quantum-classical approaches in nuclear and high-energy physics. Despite promising results in lower-dimensional datasets, the transition to higher-resolution images revealed significant obstacles, particularly in overcoming \emph{mode collapse}. 

While we introduced regularization methods, such as the KL-divergence loss, which had successfully stabilized training on small-scale images, these methods proved insufficient to prevent mode collapse in higher-dimensional settings. This suggests that the complexity of the LArTPC data, combined with the increased sparsity of non-zero-value pixels in higher resolutions, overwhelms the model's capacity to learn meaningful representations. The sparsity likely results in the model converging to simplified outputs, such as single bright spots, rather than capturing the diverse and complex particle trajectories.

The scaling of the quantum circuit to handle 8 qubits and process $24\times 32$ pixel images presented an exciting opportunity to increase the model’s generative capabilities. However, our experiments highlighted the need for more advanced regularization techniques or architectural modifications to ensure that the model can better balance the complexity of higher-resolution data without falling into mode collapse.

A potential future direction could involve incorporating more sophisticated quantum architectures or investigating alternative hybrid architectures could help address the limitations observed in the current model's performance at scale.

In summary, while our experiments demonstrate the potential of quantum-enhanced models for generative tasks in nuclear and high-energy physics, significant work remains to overcome the challenges of scaling to higher-resolution data and avoiding common pitfalls such as mode collapse. These insights serve as a foundation for further exploration of scalable quantum-classical models for subatomic physics experiments.
%-----------------------------------------------------------
\section{Conclusion}
\label{sec:conclusion}

In this study, we explored the application and scaling of quantum-enhanced generative models for high-resolution image generation in the context of LArTPC. By increasing the number of qubits and scaling the model to process larger image sizes, we aimed to capture the complexity of LArTPC data. Although regularization techniques previously successful in smaller datasets were introduced, they were unable to prevent mode collapse in higher-resolution settings, likely due to the increasing sparsity of the data. This mode collapse persisted across various training schemes and regularization methods, indicating the need for further refinement of the model's architecture and regularization techniques.

We did not investigate the effect of hardware noise in this work, as we believe this aspect is better explored in the context of diffusion models or once the mode collapse problem has been adequately addressed. Hardware noise presents a significant challenge in real-world quantum computing, but solving the fundamental issue of mode collapse remains a priority for improving generative model performance on high-resolution data. We leave the exploration of hardware noise for future work once more stable generative models are developed.

This work offers valuable insights into the limitations and potential of quantum-enhanced generative models for neutrino physics experiments, with important implications for the development of scalable models in nuclear and high-energy physics. Future research should focus on overcoming mode collapse, exploring new quantum architectures, and assessing the role of hardware noise to further advance the application of quantum computing in this domain.

%-----------------------------------------------------------
\begin{acknowledgments}
This work was partially supported by the Quantum Horizons: QIS Research and Innovation for Nuclear Science program at ORNL under FWP ERKBP91. This research was supported in part by an appointment to the RENEW Pathway Summer Schools Program at the Oak Ridge National Laboratory, sponsored by the U.S. Department of Energy and administered by the Oak Ridge Institute for Science and Education.
\end{acknowledgments}
\bibliography{ref}
\appendix

\section{Dataset complexity}
\label{sec:datacomplexity}

To emphasize the complexity of the LArTPC dataset compared to the simpler MNIST dataset, and to illustrate why LArTPC poses a greater challenge for model training, the entropy distribution, fractal dimension, and PCA visualization are plotted. In terms of entropy, a complex dataset like LArTPC is expected to exhibit a broader and more uniform distribution, as it contains diverse and variable image structures, leading to higher and more varied information content. For the fractal dimension, a more complex dataset should show higher values, reflecting greater structural complexity within the images, which corresponds to the characteristics of real-world data. 

\begin{figure}[h]
\centering
\includegraphics[width=\linewidth]{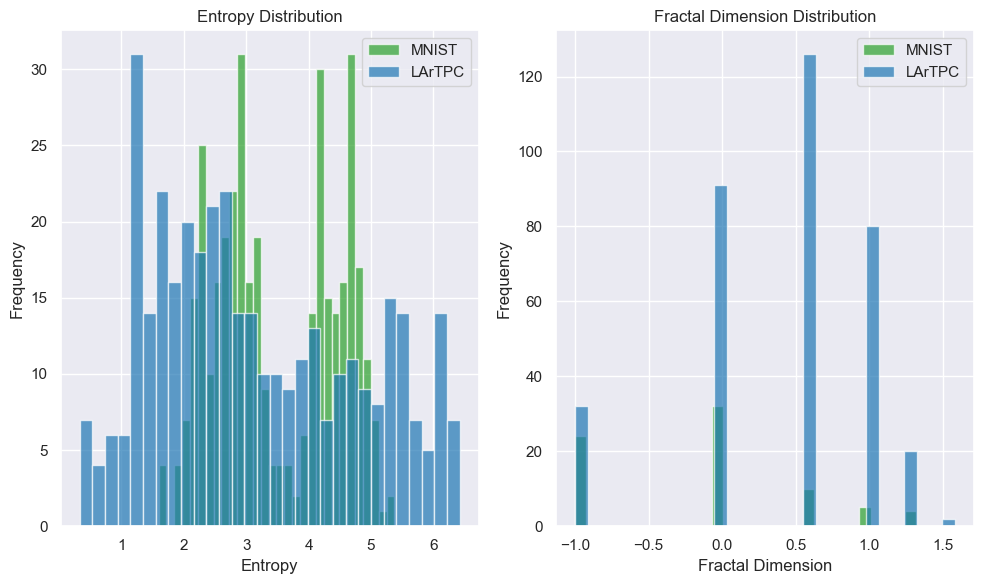}\\
\includegraphics[width=\linewidth]{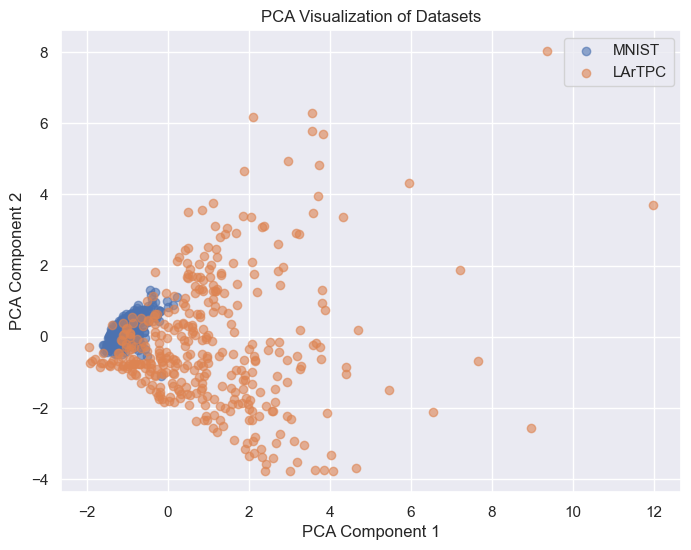}\\
\caption{Comparison of dataset complexity between LArTPC and MNIST using entropy distribution, fractal dimension, and PCA visualization. The top row shows the broader entropy and higher fractal dimension of the LArTPC dataset, indicating its greater structural complexity compared to MNIST. The bottom row presents the PCA visualization, where LArTPC exhibits wider spread across components, reflecting higher feature variance and illustrating the increased challenge of training models on LArTPC data.}
\label{fig:complexityplots}
\end{figure}

In contrast, simpler datasets like MNIST are expected to show concentrated peaks at lower values for these metrics, indicating homogeneity and simplicity. Lastly, in the PCA visualization, a complex dataset should spread more widely across multiple components, indicating high variance in the features extracted by the model, making the learning process more challenging (See Fig.~\ref{fig:complexityplots}).

For this study, a subset of the \texttt{MNIST} dataset~\cite{MNIST} is used, focusing on the digits '0' and '1', as they were utilized in the study referenced in Ref.~\cite{Zhang_2024sfa}. The \texttt{MNIST} dataset consists of $28\times 28$ grayscale images of handwritten digits from 0 to 9.

%----------------------------------------------------------
% Begin figure with full page width in two-column layout
%----------------------------------------------------------

\section{Role of the Jacobian Determinant in Measuring Expressiveness}
\label{sec:jacobian}

In classical neural networks, particularly normalizing flows, the expressiveness of the model is tied to its ability to transform a simple base distribution into a more complex target distribution. The Jacobian determinant, $\det J$, quantifies how the transformation $f_{\theta}: \mathbb{R}^{n}\rightarrow\mathbb{R}^{n}$ changes the volume of the input space:

\begin{equation}
    \det J = \left| \frac{\partial f_\theta(x)}{\partial x} \right|,
\end{equation}

If $\det J >1$, the transformation expands the volume in the input space, whereas if $\det J <1$, it contracts the volume. A wide range of possible values for $\det J$ indicates that the model can flexibly reshape the input distribution, which is crucial for modeling complex data distributions. Therefore, the Jacobian determinant acts as a measure of the network's capacity to perform complex mappings suggesting high expressiveness when it can vary widely.

Quantum neural networks (QNNs) leverage quantum properties such as superposition, entanglement and inference. The expressiveness of a QNN is often linked to its ability to represent complex quantum states and perform transformations that classical models cannot easily replicate. However, quantum circuits are represented by unitary matrices $U$, where $UU^{\dagger} = I$, and the determinant of any unitary matrix is constrained to be on the complex unit circle $|\det U| = 1$. This constraint implies that quantum transformations are volume-preserving in the complex Hilbert space. Consequently, the Jacobian determinant for a purely QNN is always of magnitude 1 and does not provide a useful measure of expressiveness. In this case, the expressiveness comes from leveraging quantum phenomena like entanglement, which the Jacobian determinant does not capture.

For a hybrid quantum-classical model, where both classical neural network components and quantum circuits are used together, the situation becomes more interesting. The overall transformation $f_{\theta}$ consists of both classical transformations $f_{classical}: \mathbb{R}^{n}\rightarrow \mathbb{R}^{n}$ and quantum transformations $U: \mathcal{H}\rightarrow \mathcal{H}$, where $\mathcal{H}$ is a Hilbert space. Nonetheless, as demonstrated in Ref.~\cite{Zhang_2024sfa}, the overall determinant of the hybrid network only relies on the classical component since the quantum part is unitary. Thus, while the determinant of the Jacobian, $\det J$, can offer insights into the expressiveness of the classical components of the hybrid quantum-classical network, it is not a complete measure for the entire model expressiveness. A more comprehensive assessment would combine the Jacobian determinant with metrics that specifically account for the quantum part's expressiveness, such as the Quantum Fisher Information (QFI). However, calculating the QFI for a parameterized quantum circuit can be computationally expensive, especially as the number of qubits and circuit depth increases. This added complexity could slow down the training process or require more advance hardware. Furthermore, the usefulness of incorporating QFI depends on the specific task. For tasks that require significant quantum advantage, such as those involving quantum state discrimination or entanglement detection, QFI might be highly beneficial. For more classical tasks, the benefit may be less pronounced.

\section{KL-Divergence Regularization}
\label{sec:kldivergence}
In the context of neural networks, especially in Variational Autoencoders (VAEs) or latent variable models, $\mu$ (mean) and $\sigma$ (standard deviation) represent the parameters of a Gaussian (normal) distribution in the latent space. 

The encoder network maps the input data to a lower-dimensional latent space, and instead of directly outputting a fixed latent vector for each input, the encoder outputs a distribution over the latent space, typically modeled as a Gaussian distribution. This distribution is parameterized by the mean $\mu$ and the log variance $\log(\sigma^2)$, where $\sigma$ is the standard deviation.

The neural network outputs the mean $\mu$ of the latent Gaussian distribution for each input sample, representing the center or `location' of the distribution in the latent space, determining where the model expects the latent variable to be most likely. Similarly, the network outputs the log variance $\log(\sigma^2)$, which is used to compute the standard deviation $\sigma$ of the latent distribution. The standard deviation controls the spread or uncertainty around the mean, indicating how confident the network is about the location of the latent variable.

To make the sampling process differentiable, the reparameterization trick is applied. Instead of directly sampling from the distribution parameterized by $\mu$ and $\sigma$, the latent variable $z$ is computed as:

\begin{equation}
    z = \mu + \sigma \cdot \epsilon,
\end{equation}

where $\epsilon \sim \mathcal{N}(0, I)$ is a random variable sampled from a standard normal distribution. This trick allows the neural network to backpropagate through the sampling process, maintaining the differentiability of the entire model.

In this context, $\mu$ represents the expected value or `most likely' latent representation for a given input, while $\sigma$ controls the variability around $\mu$, allowing the model to capture a range of possible latent variables for each input. This variability enables the model to generalize better and explore the latent space more effectively, thereby reducing problems such as mode collapse.

\end{document}